\documentclass{aastex63}
\usepackage{amsmath}

\submitjournal{ApJ}

\shorttitle{Alert follow-up by IACTs}
\shortauthors{Fiorillo et al.}

\graphicspath{{./}{/}}

\begin{document}

\title{A test of the hadronic origin of $\gamma$-rays from blazars with up to month-later follow-up of IceCube Alerts with Imaging Air Cherenkov Telescopes}

\correspondingauthor{Damiano F. G. Fiorillo}
\email{dfgfiorillo@na.infn.it}

\author{Damiano F. G. Fiorillo}
\affiliation{Dipartimento di Fisica "Ettore Pancini", Universitá degli studi di Napoli "Federico II", Complesso Univ. Monte S. Angelo, I-80126 Napoli, Italy}
\affiliation{INFN - Sezione di Napoli, Complesso Univ. Monte S. Angelo, I-80126 Napoli, Italy}

\author[0000-0002-7669-266X]{Konstancja Satalecka}
\affiliation{Deutsches Elektronen-Synchrotron (DESY), D-15738 Zeuthen, Germany}

\author[0000-0003-3509-3457]{Ignacio Taboada}
\affiliation{School of Physics, Georgia Institute of Technology\\387 State St.\\Atlanta GA, 30332. USA.}

\author{Chun Fai Tung}
\affiliation{School of Physics, Georgia Institute of Technology\\387 State St.\\Atlanta GA, 30332. USA.}

\begin{abstract}
The sources of IceCube neutrinos are as yet unknown. The multi-messenger observation of their emission in $\gamma$-rays can be a guide to their identification, as exemplified by the case of TXS 0506+056. We suggest a new method of searching for $\gamma$-rays with Imaging Air Cherenkov Telescopes from sources in coincidence with possible astrophysical neutrinos. We propose that searches of $\gamma$-rays are extended, from the current practice of only a few days, to up to one month from a neutrino alert. We test this strategy on simulated sources modeled after the blazar \emph{TXS 0506+056-like}, emitting neutrinos and $\gamma$-rays via photohadronic interactions: the $\gamma$-rays are subsequently reprocessed in the VHE range. Using MAGIC as a benchmark example, we show that current Cherenkov Telescopes should be able to detect$\gamma$-ray counterparts to neutrino alerts with a rate of approximately one per year. It has been proposed that the high-energy diffuse neutrino flux can be explained by $\sim$ 5\% of all blazars flaring in neutrinos once every 10 years, with a neutrino luminosity similar to that of TXS 0506+056 during the 2014-2015 neutrino flare. The implementation of our strategy could lead, over a timescale of one or few years, either to the detection of this subclass of blazars contributing to the diffuse neutrino flux, or to a constraint on this model.
\end{abstract}

\keywords{Neutrino astronomy, neutrino telescopes, gamma-ray astronomy, gamma-ray telescopes, gamma-ray transient sources, blazars}

\section{Introduction}
\label{sec:intro}

Neutrinos and very high energy (VHE, E $>$ 100 GeV) $\gamma$-rays are produced together in hadronic interactions whenever cosmic rays (CRs) encounter matter or photon fields inside a cosmic source or in its surroundings.
Observations of these neutral messengers give us important clues in determining the origins of CRs. The IceCube neutrino observatory has discovered a diffuse astrophysical flux of neutrinos in the 10 TeV to 10 PeV energy range \citep{2014PhRvL.113j1101A,2016ApJ...833....3A,PhysRevLett.125.121104}. The flux is isotropic over the sky, evidence of extra-galactic origin. The flux is consistent with an equal flux of neutrinos for each flavor as expected from standard neutrino oscillations over astrophysical baselines \citep{2015PhRvL.114q1102A}. An angular uncertainty of, at best, 0.5$^\circ$ for neutrino directions in IceCube has prevented the clear association of the astrophysical diffuse neutrino flux with a class or classes of astrophysical objects.
The angular resolution of a typical ground-based $\gamma$-ray telescope (imaging atmospheric Cherenkov telescope, IACT) is $\sim$0.06$^{\circ}$, which makes sources identification much easier than with neutrino telescopes. 
Although neutrinos can only originate from hadronic processes, $\gamma$-rays can as well be produced in leptonic interactions. They also suffer from absorption on low energy photon background, the Extragalactic Background Light (EBL) \citep{2013APh....43..112D}. At the energy of 100\,TeV (10$^{14}$\,eV), their horizon is limited to the Virgo Cluster at $\sim$16 Mpc (65 million light-years) away. 
Neutrinos, on the other hand, can travel unimpeded through cosmological distances and penetrate very dense environments. 
Therefore only by combining VHE $\gamma$-ray and neutrino observations can we resolve the origins of CRs.

Neutrinos of likely astrophysical origin are announced in realtime, with a typical latency of $\sim$30~s, by IceCube. The methodology of alerts was updated on June 2019 \citep{2019ICRC...36.1021B}. Track-like neutrinos, dominated by muon neutrinos, are reported as \textit{Gold} alerts if their average signalness is 50\%, i.e., in a large set of Gold events, half are of astrophysical origin. \textit{Bronze} alerts have an average signalness of 30\%. Realtime alerts have a localization uncertainty of $\sim1^\circ$, which enables follow-up by IACTs, with typical camera diameter spanning 3.5-5$^{\circ}$.

Here we propose a new strategy for searching for VHE $\gamma$-ray flares in coincidence with likely astrophysical neutrinos. This new method can test if, as suggested by modelers, a small fraction of blazars are responsible for the astrophysical high-energy neutrino diffuse flux. This test can be conclusive in a campaign that is as short as one year, in which IACT follow-up $\sim$6 likely astrophysical neutrinos. We propose that IACTs observe neutrino alerts with a time delay of up to one month.

In \citet{2017ApJ...835...45A} the IceCube collaboration shows that gamma-ray bright blazars can contribute to no more than $\sim$27\% of the diffuse flux.
So, it is somewhat puzzling that TXS~0506+056, one of the brightest$\gamma$-ray blazars, has been identified as a candidate neutrino source.
On September 2017, IceCube reported neutrino alert IceCube-170922A \citep{2017GCN.21916....1K}. Both $Fermi$/LAT with GeV gamma rays \citep{LATATEL} and MAGIC with $\gtrsim$100~GeV gamma rays \citep{MAGICATEL} reported TXS~0506+056 to be flaring at the time. A large number of instruments followed these observations, from radio to VHE gamma rays. The accidental coincidence of the neutrino with$\gamma$-rays was ruled out at 3$\sigma$ \citep{0506science1}. 
An archival review of IceCube data in the direction of TXS~0506+056 revealed that it had flared in neutrinos 
for 110 days in 2014-2015, assuming a box time-profile.
A background fluctuation was ruled out as the flare's origin with 3.5$\sigma$ \citep{0506science2}. Interestingly, $Fermi$/LAT found no evidence for an enhanced GeV$\gamma$-ray emission during this time \citep{2019ApJ...880..103G,2018MNRAS.480..192P}.

Modeling the 2017 neutrino-gamma-ray coincidence has been done by multiple authors \citep{Cerruti19, Gao19, MAGIC_TXS_paper2, Keivani18, Petropoulou20}. However, modeling the 2014-2015 gamma-ray-dark neutrino flare has proven a theoretical challenge. It might suggest that the neutrino and photon emitting regions are separated, and the first one is opaque to photons \citep{Reimer19, Xue19}. Another solution is to channel the photon emission into the MeV band, where the upper limits are less constraining \citep{Rodrigues19}.

\citep{Halzen_2019} have provided a phenomenological model, henceforth the HKWW model, of the 2014-2015 TXS~0506+056 neutrino flare. The authors take into account the constraints on the gamma-ray blazars contribution to the diffuse neutrino flux. They argue that a subclass of blazars, about 5\% of all blazars, each flaring once every $\sim$10 years, are responsible for the all-sky high-energy neutrino flux. HKWW assumes that neutrinos and$\gamma$-rays are produced via the p-$\gamma$ process with a power-law spectrum of index -2.2. After internal re-processing and EBL attenuation, a \emph{TXS-0506+056-like} $\gamma$-ray flare, in temporal coincidence with neutrinos, would peak at $\sim$100~GeV. This energy is well accessible to ground-based Cherenkov telescopes. In what follows, we investigate if and how the HKWW scenario could be directly tested by realtime follow-up of neutrino alerts with the currently operating IACTs. 

Despite the successful follow-up campaign of IceCube-170922A, attempts of the current generation IACTs to follow up other IceCube alerts or neutrino events of likely astrophysical origin \citep{Aartsen:2016qbu,2017ICRC...35..618S,2019ICRC...36..787S,2019ICRC...36..633B,2019ICRC...36..782S} have given no results.
The publications cited above report typical follow-up strategies among IACTs. In general, they are very similar: all of them implemented fast, automatic re-pointing which allow an immediate observation, if the alert arrives during a dark night. Sporadically, observations are taken with up to a few (2-3) days after the alert, in case of a full Moon or bad weather. Typical exposure ranges from 0.5\,h to 3\,h.

This strategy is well suited for transient sources, emitting on the time scales of hours to days. However, it might lead to a situation where the alert is not observed if not immediately visible by the telescope. This excludes a possibility to test models predicting longer emission duration.
In this work, we propose an observational test of the HKWW model using IACTs. As an example of the current generation IACT system, we will use the MAGIC telescopes since their performance is documented in the literature with the most detail \citep{2016APh....72...76A}.
We will show that current IACTs should detect a $\gamma$-ray counterpart from a \textit{TXS~0506+056-like} flaring blazar with a rate of 1.0--1.5/year if they perform an IceCube alert follow-up with a delay of up to one month.  

\section{Simulation of IceCube Alerts}

We conduct simulations of populations of neutrino flaring blazars that follow the HKWW model with python software FIRESONG \citep{FIRESONG_JOSS}. To generate the neutrino fluxes, FIRESONG considers the effect of cosmic expansion and is capable of simulating different source populations characterized by evolution models, local number densities, and emission profiles.

The total number of the neutrino flaring blazars in the universe, $N_\mathrm{total}$, is given by
\begin{equation}
    N_\mathrm{total} = \int^{z_{max}}_0 \rho(z) \frac{dV_c}{dz} dz\,,
\end{equation}
where $\rho(z)$ is the number density as a function of redshift $z$, $z_{max}$ is the farthest redshift considered, and $dV_c$ is the differential comoving volume.
We have assumed that blazars do no evolve in density or luminosity, so $\rho(z)$ is the same as the local density over the entire universe. Following HKWW, we assume that the local density for all blazars, neutrino flaring or not, is  $1.5\times 10^{-8}$~Mpc$^{-3}$, and we consider that 1\%, 5\% or 10\% of all blazars (local densities of $1.5\times 10^{-10}$~Mpc$^{-3}$, $7.5\times 10^{-10}$~Mpc$^{-3}$, $1.5\times 10^{-9}$~Mpc$^{-3}$ respectively) produce flares that result in neutrinos.

We assume that blazar flares are standard candles, i.e., all blazar flares have the same rate of neutrino production per unit of energy in their rest frame, $Q'_\nu(E'_\nu)=\frac{dN_\nu}{dE'_\nu dt'}$, as TXS~0506+056; here the primed quantities are in the blazar rest frame.
This is equivalent to stating that all neutrino blazar flares have the same neutrino luminosity in their rest frame $L'_\nu = \int E'_\nu Q'_\nu(E'_\nu) dE'_\nu$. The true luminosity function of neutrino sources is likely to be significantly more complicated, but as described in \citet{2016PhRvD..94j3006M}, the use of a characteristic luminosity is appropriate. 

FIRESONG simulations have been conducted so that the cumulative neutrino flux from all flaring blazars matches the muon neutrino diffuse flux:
\begin{equation}
\label{eq:diffuseNuMu}
\frac{d\Phi_{\nu+\bar{\nu}}}{dE_\nu} = 1.01 
\left( \frac{E}{100\mathrm{TeV}} \right)^{-2.19} \cdot 10^{-18} \mathrm{GeV}^{-1}\mathrm{cm}^{-2}\mathrm{s}^{-1}\mathrm{sr}^{-1}.
\end{equation}
Under this assumption, the product of neutrino luminosity (during the flare) and local density is a constant that depends on the neutrino diffuse flux spectrum.
The best fit spectral index by IceCube for the 2014-2015 TXS~0506+056 neutrino flare is -2.2; this is the spectral index also used by HKWW. However, we use the value -2.19 in FIRESONG simulations since the published IceCube's Gold alert rates assume this index and the diffuse flux in equation~\ref{eq:diffuseNuMu} also has this best fit spectral index. As a check of our calculations, we have verified that FIRESONG simulations of a fraction of blazars flaring once every 10 years saturate the diffuse flux; in fact, we obtain the fraction of $\sim$3\%. 

For a given simulated neutrino flare at luminosity distance $d_L$, the spectrum is:
\begin{equation}
   \frac{d\Phi_{\nu+\bar{\nu}}}{dE_\nu} = (1+z)^2\frac{Q'_{\nu}\left[E_\nu (1+z)\right]}{4\pi d_L^2}=A_\nu \left(\frac{E_\nu}{E_0}\right)^{-\gamma},
\end{equation}
where $\gamma = 2.19$ is spectral index of the flare and $d_L$ is the redshift-dependent luminosity distance. Finally, we assume that neutrino flares last 82 days, box shaped, in the local reference frame. This matches the observation of the TXS~0506+056 flare, with redshift z=0.3365, lasting 110 days.

\begin{table}[t]
    \centering
    \begin{tabular}{l|rrr}
    \hline
          & North & Horizon & South \\
         \hline
        Astrophysical Gold alert rate (1/yr) & 0.95 & 3.89 & 1.00  \\
        \hline
    \end{tabular}
    \caption{Simulated Gold alert rate from astrophysical neutrino sources corresponding to the diffuse flux in equation~\ref{eq:diffuseNuMu}. Because the set of Gold alerts has a probability of 50\% of being astrophysical in origin, the total predicted Gold alert rate, including background, can be computed simply by multiplying table numbers by two.}
    \label{tab:alertrates}
\end{table}

A given simulation run results in a list of flaring blazars, each characterized by a redshift, right ascension, declination, and the average muon neutrino spectrum at Earth during the flare. To calculate the response of IceCube to these simulations, we use the effective areas for IceCube Gold alerts with coarsely-binned declination dependence, as given in \citet{2019ICRC...36.1021B}. These effective areas are averages provided for the "southern" sky, $\delta<-5^\circ$, near the celestial equator, $-5^\circ < \delta < 30^\circ$ and in the "northern" sky, $\delta>30^\circ$. The effective area is largest near the celestial equator and smallest in the southern sky. 

We have also calculated the rate of Gold alerts produced by astrophysical sources using the spectrum on equation~\ref{eq:diffuseNuMu}. We obtain 5.8 alerts/year, which compares well with IceCube's prediction of 6.6  astrophysical alerts/year \citep{2019ICRC...36.1021B}. The difference is probably due to \citet{2019ICRC...36.1021B} using neutrino effective areas more finely binned in declination, which are not publicly available. It should be noted that the number of alerts per declination band depends on the diffuse flux and not on the HKWW model. The Gold alert rates above do not include atmospheric neutrino and muon background, which for Gold alerts, doubles the rate of alerts. We have not simulated background; instead, we multiply by two the rate at each declination band. Gold alert rates from astrophysical sources are displayed in Table \ref{tab:alertrates}.

\section{Simulation of VHE gamma rays from neutrino alerts}\label{sec:gammaraysalerts}

In this section we describe how we compute the $\gamma$-ray flux associated with each simulated IceCube alert. 

Since neutrinos and $\gamma$-rays are produced in $p\gamma$ processes within the astrophysical source, the spectra with which they are injected at the source are connected.
Therefore, assuming flavor equipartition,
\begin{equation}\label{eq:gammainjected}
    (1+z)^2 \frac{Q'_\gamma\left[E_\gamma(1+z)\right]}{4\pi d_L^2}=A_\nu \left(\frac{E_\gamma}{2E_0}\right)^{-\gamma}.
\end{equation}

Gamma-rays are injected with the spectrum in equation~\ref{eq:gammainjected}, but are subsequently reprocessed by $\gamma\gamma$ collisions, according to the discussion by HKWW. 
Since a full description of the reprocessing would be complicated and rather model-dependent, we adopt the same phenomenological approach and express the spectrum after internal reprocessing as
\begin{equation}\label{eq:reprocessed}
    E_\gamma^{'2} Q^{'\text{rep}}_\gamma(E'_\gamma)=\mathcal{N} e^{-\frac{E'_L}{E'_\gamma}-\frac{E'_\gamma}{E'_H}},
\end{equation}
where $z$ is the redshift of the source, $E'_L=134$ GeV and $E'_H=27$ TeV are the low and high energy cutoff in the source rest frame. This corresponds to 100 GeV and 20 TeV in the Earth's frame for the redshift of TXS 0506+056, namely the values used in the HKWW model. The normalization constant is chosen to match the total power emitted in the reprocessed (equation~\ref{eq:reprocessed}) and the injected (equation~\ref{eq:gammainjected}) spectra. In calculating the power from equation~\ref{eq:gammainjected}, we have taken into account that IceCube's observations are done only over a limited energy, i.e., there are hard lower and upper bounds to the gamma-ray power-law spectra between 30 TeV and 3 PeV.

After imposing this matching, the normalization of the reprocessed spectrum can be expressed in terms of the normalization of equation~\ref{eq:gammainjected}: the final result for the $\gamma$-ray flux at Earth is
\begin{equation}\label{eq:gammarayflux}
    \frac{d\phi_\gamma}{dE_\gamma}=4.2 A_\nu \left(\frac{E_\gamma}{E_0}\right)^{-2} \exp\left[-\frac{E'_L}{E_\gamma(1+z)}-\frac{E_\gamma(1+z)}{E'_H}\right]e^{-\tau_{\gamma\gamma}(E_\gamma,z)}.
\end{equation}

The exponential factor $e^{-\tau_{\gamma\gamma}(E_\gamma,z)}$ accounts for the suppression caused by the interaction with the EBL, so that $\tau_{\gamma\gamma}(E_\gamma,z)$ is the optical depth of $\gamma\gamma$ interaction for a gamma ray with energy $E_\gamma$ originated at redshift $z$. We adopt the EBL model from~\citet{Gilmore:2011ks}.

\section{Detectability of IceCube Alerts}

The detectability of each IceCube alert depends on the response of the $\gamma$-ray telescope, in our example, MAGIC. The aim of this section is to show how we model the telescope response.
We use the instrument response functions (IRFs) provided in~\citet{2016APh....72...76A}. The effective areas and the background rates for the MAGIC telescopes are given for two zenith bins, namely $\theta<30^{\circ}$ and $30^{\circ}<\theta<45^{\circ}$, with $\theta$ the zenith angle of the source. We do not consider observations with larger zenith angles. For each zenith bin the number of expected events for an observation time $T_\text{obs}$ is:
\begin{equation}
    N_\text{signal}=\int_{100\;\text{GeV}}^{30\;\text{TeV}} dE_\gamma \frac{d\phi_\gamma}{dE_\gamma} A_\text{eff} (E_\gamma,\theta) T_\text{obs},
\end{equation}
where $A_\text{eff} (E_\gamma,\theta)$ is the effective area in the corresponding energy bin (the angular dependence is related to the choice of the zenith bin) and $\frac{d\phi_\gamma}{dE_\gamma}$ is the photon flux per unit energy as defined in equation~\ref{eq:gammarayflux}. For the results shown in this work, we conservatively use an observation time $T_\text{obs}=0.5$ h; we discuss in Section~\ref{sec:conclusions} the effects of a different choice. Finally, we impose a lower energy threshold for the analysis of 100 GeV.

The number of signal events is to be compared with the number of background events $N_\text{background}$ over the entire energy range, as given by the MAGIC Collaboration\footnote{For reference, we find that over the entire energy range the expected number of background events is $92.2$ and $117.3$ for the two zenith bins from $0^\circ$ to $30^\circ$ and from $30^\circ$ and $45^\circ$ respectively, assuming half an hour of observation time.}. We use the procedure proposed in equation~5 of \citet{Li:1983fv} and evaluate the significance (number of standard deviations) for detection as
\begin{equation}\label{eq:significancelima}
    S=\frac{N_\text{signal}}{\sqrt{N_\text{signal}+(1+\alpha) N_\text{background}}}.
\end{equation}
Here $\alpha=T_\text{obs}/T_\text{off}$ is the ratio between the exposure time of the experiment $T_\text{obs}$ and the exposure time for the estimation of the background rate $T_\text{off}$. For this work we conservatively take $\alpha=1$. This assumption is conservative because in principle $T_\text{off}$ could be much larger than $T_\text{obs}$, leading to smaller values of $\alpha$ and larger significance. Therefore, with this assumption we are slightly underestimating the potential of the observation. Using equation~\ref{eq:significancelima}, we can compute the significance for detection of each alert from the list simulated with FIRESONG. There are two complications to this scenario which need to be accounted for. The first one is the flare duration: are we going to catch the $\gamma$-ray emission in time? The second one is the visibility of the source, i.e. dependence of the telescope effective area on the zenith angle and astronomical constraints that can prevent the observation. We are now going to describe how we deal with both these complications.

\subsection{Time constraints: flare duration and follow-up delay}

As discussed in Section~\ref{sec:gammaraysalerts}, we assume that each alert is connected with a flare whose duration in the source rest frame is $T'_\text{flare}=82$ days. The neutrino alert can happen anytime during this time window. For this reason, the flare may be over, by the time that an IACT has managed to make an observation. 
Therefore whether a source can be detected or not depends on when the observation is performed. Let $T_\text{delay}$ be the interval between the time when IceCube sent the alert and the time at which the IACT performs the measurement. This value depends on many external factors, 
e.g. presence of the Moon or weather at the site. For definiteness, we assume $T_\text{delay}$ to be randomly distributed with a maximum value of $T_\text{max}=30$ days. Our aim is to find the probability that the experiment can perform the observation before the end of the flare. If we measure the time starting from the beginning of the flare, and we let $T_\text{i}$ be the time at which IceCube observed the neutrino and sent the alert, then this requirement can be expressed as
\begin{equation}\label{eq:probtime}
    T_\text{i}+T_\text{delay}<T'_\text{flare}(1+z),
\end{equation}
where we have taken into account the dilation of the flare duration due to redshifting. We assume $T_\text{i}$ to be uniformly distributed within the flare duration, namely between $0$ and $T'_\text{flare}(1+z)$. The probability that equation \ref{eq:probtime} is verified is then obtained integrating over the distributions for both $T_\text{i}$ and $T_\text{delay}$ and is equal to
\begin{equation}\label{eq:distributionflare}
    P_\text{meas}(z)=1-\frac{T_\text{max}}{2T'_\text{flare}(1+z)}.
\end{equation}
This is the probability that an alert at redshift $z$ is observable by an IACT,  while the flare is still active.

\subsection{Observability constraints: source's ephemeris, Sun and Moon}

The second complication to be accounted for is the zenith angle under which an IACT can detect the alerts. As mentioned above, the MAGIC Collaboration provides two different sets of IRFs for two different ranges of the zenith angle with which the source is observed. This zenith angle depends on the declination of the alert, to be compared with the latitude of the experiment. It also depends on the time of the year at which the alert is sent. If the position of the alert is too close to the Sun's position during the relevant time of the year, the observation might be impossible to perform. To account for this dependence, we want to determine the probability that the source is visible, without Sun or Moon interference, with a zenith angle between $0^\circ$ and $30^\circ$ ($P_{0^\circ-30^\circ}(\delta)$), between $30^\circ$ and $45^\circ$ ($P_{30^\circ-45^\circ}(\delta)$), and between $45^\circ$ and $180^\circ$ ($P_{45^\circ-180^\circ}(\delta)$, i.e., unobservable) for a random time of the alert during the year. These probabilities will depend upon the source declination.

To determine this dependence, we have simulated the position of the Sun and the Moon as a function of the time over 5 years using the Python package \texttt{PyEphem} (\cite{2011ascl.soft12014R}). The procedure we adopt for a fixed alert declination and right ascension is the following:
\begin{enumerate}
\item For a given initial time of the alert, we determine the best (minimum) zenith angle with which the source can be seen within 30 days from the alert, under the condition of moonless nights. For the latter condition we require that the Sun is at least $18^\circ$ below the horizon (so the zenith angle of the Sun is larger than $108^\circ$) and the center of the Moon is $0.25^\circ$ below the horizon (so the zenith angle of the Moon is larger than $90.25^\circ$);
\item Repeating the procedure for all the initial alert times over 5 years we obtain the best zenith angle $\tilde{\theta}(t)$ that can be achieved for observation as a function of the initial alert time. The probability that the alert is seen in a certain bin of zenith angles between $\theta_i$ and $\theta_j$ can be written as
\begin{equation}
    P_{\theta_i - \theta_j}(\delta)=\frac{\int dt \Theta\left[\tilde{\theta}(t,\delta)-\theta_i\right]\Theta\left[-\tilde{\theta}(t,\delta)+\theta_j\right]}{\int dt},
\end{equation}
where $\Theta$ is the Heaviside function and the integral is performed over the 5 years on which we have done the simulation.
\end{enumerate}

The output of this procedure is the probability that the source is visible by MAGIC within each of the three zenith bins, for a random alert time. As we have averaged arrival time of the alerts and the ephemerides for the Moon and the Sun over a period of 5 years, we arrive at a probability that depends only on the alert declination. The dependence on the right ascension drops out when we average over all possible alert times because of the periodicity of the Sun's motion.

\begin{figure}
    \centering
    \includegraphics[width=0.7\textwidth]{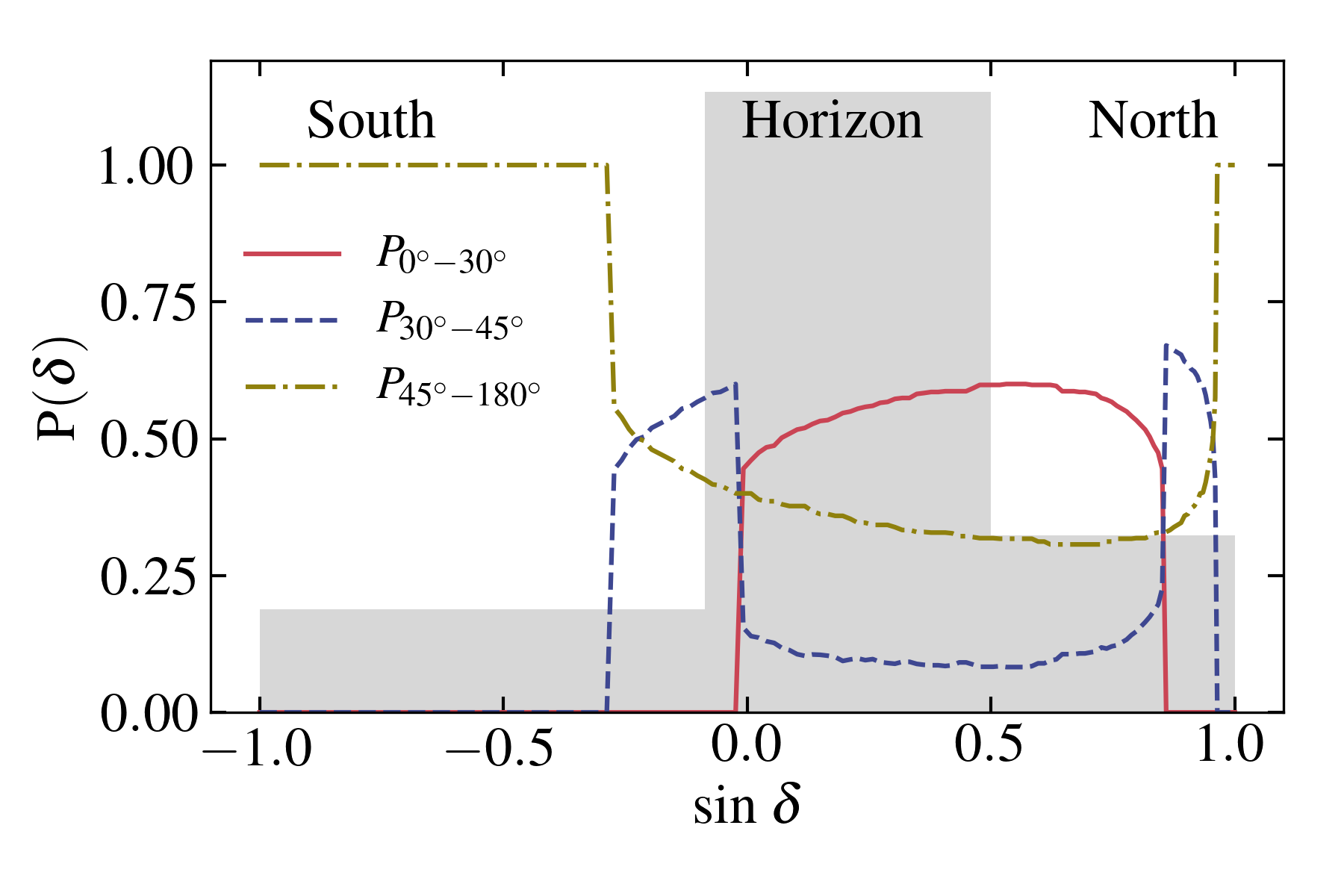}
    \caption{Probability of source observation at different zenith angles as a function of the source declination. We show the probability that the source is observed with a zenith angle between $0^\circ$ and $30^\circ$ (red solid), between $30^\circ$ and $45^\circ$ (blue dashed) and between $45^\circ$ and $180^\circ$ (golden dotted-dashed) for a random alert time as a function of the sine of the source declination. The grey histogram is the angular distribution of the alerts considered in this work, normalized to unit area.}
    \label{fig:VisibilityAlerts}
\end{figure}

We show the probability of observing the source in each of the three zenith bins as a function of the source declination in Figure~\ref{fig:VisibilityAlerts}. Since the latitude of MAGIC is approximately $28^\circ$, the source can be observed with a zenith angle smaller than $30^\circ$ only for declinations between $28^\circ\pm 30^\circ$; in the same way it can be observed with a zenith angle smaller than $45^\circ$ only for declinations between $28^\circ\pm 45^\circ$. For this reason the solid red and dashed blue curve in Figure~\ref{fig:VisibilityAlerts} are non-vanishing only in the corresponding intervals. We also show the distribution of the alerts considered in this work, highlighting the different regions of declination of southern hemisphere, horizon and northern hemisphere. 
The probabilities in Figure~\ref{fig:VisibilityAlerts} are used to weight the observations of the alerts according to the different IRFs provided by the MAGIC Collaboration for each zenith bin. It should be noted that most of IceCube's alerts are in declinations $\delta=-5^\circ$, to $\delta=30^\circ$, which is well matched to telescopes at latitudes of $\sim30^\circ$, such as MAGIC or VERITAS. 

\subsection{Results}\label{sec:results}

Using our prediction of 11.6 Gold alerts per year, half astrophysical and half due to background, and using the constraint imposed by the Sun and the Moon for following up alerts with a maximum delay, $T_{max}$, of 30 days, we find that MAGIC should be able to observe (follow-up) 6.4 Gold Alerts per year.

Using both the time duration constrains and the constrains imposed by the Moon and the Sun, we adopt the following procedure to determine how many alert follow-ups per year result in a detection:
\begin{enumerate}
    \item For each simulated alert, defined by its redshift $z^i$, its declination $\delta^i$ and the normalization of the neutrino flux $A^i_\nu$, we compute the significance of detection according to equation~\ref{eq:significancelima}   using the IRFs for both zenith bins according to the criterion of 5 standard deviations of significance. We denote the two significances as $S^i_{0^\circ-30^\circ}$ and $S^i_{30^\circ-45^\circ}$;
    \item In order to take into account the visibility of the source, we assign to each alert a weight obtained by averaging over the probabilities in Figure~\ref{fig:VisibilityAlerts}. Our threshold for detection is $5$ standard deviations, so the weight assigned to each alert is
    \begin{equation}\label{eq:weightalerts}
      W^i=P_{0^\circ-30^\circ}(\delta^i) \Theta\left(S^i_{0^\circ-30^\circ}-5\right)+P_{30^\circ-45^\circ}(\delta^i) \Theta\left(S^i_{30^\circ-45^\circ}-5\right).
    \end{equation}
    This is the probability that the alert is observed at a random initial time assuming an infinite flare duration;
    \item Finally, to account for the finite duration of the flare, we multiply the weights by the probability in equation~\ref{eq:distributionflare} for the corresponding alert redshift.
\end{enumerate}

\begin{figure}
    \centering
    \includegraphics[width=0.3\textwidth]{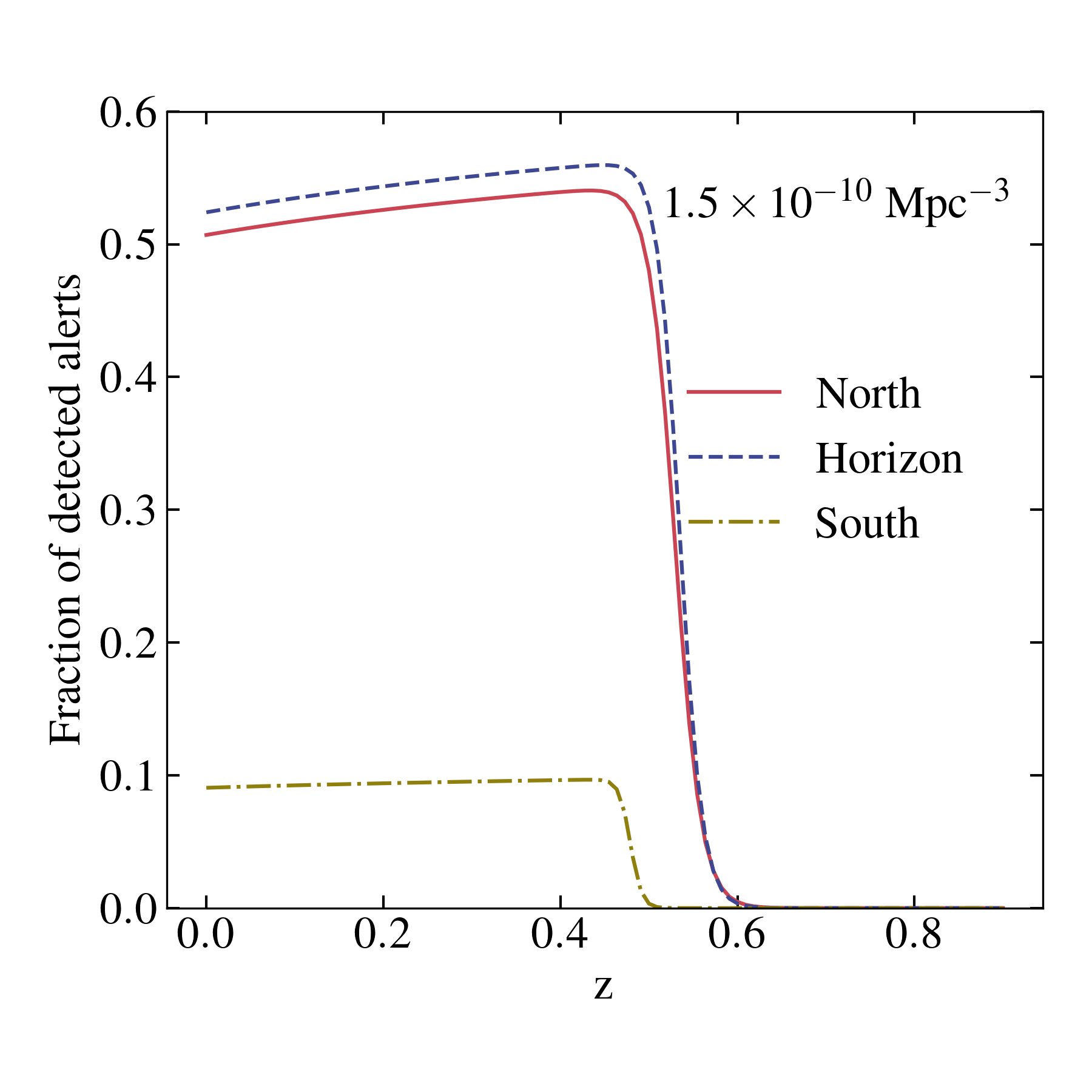}
    \includegraphics[width=0.3\textwidth]{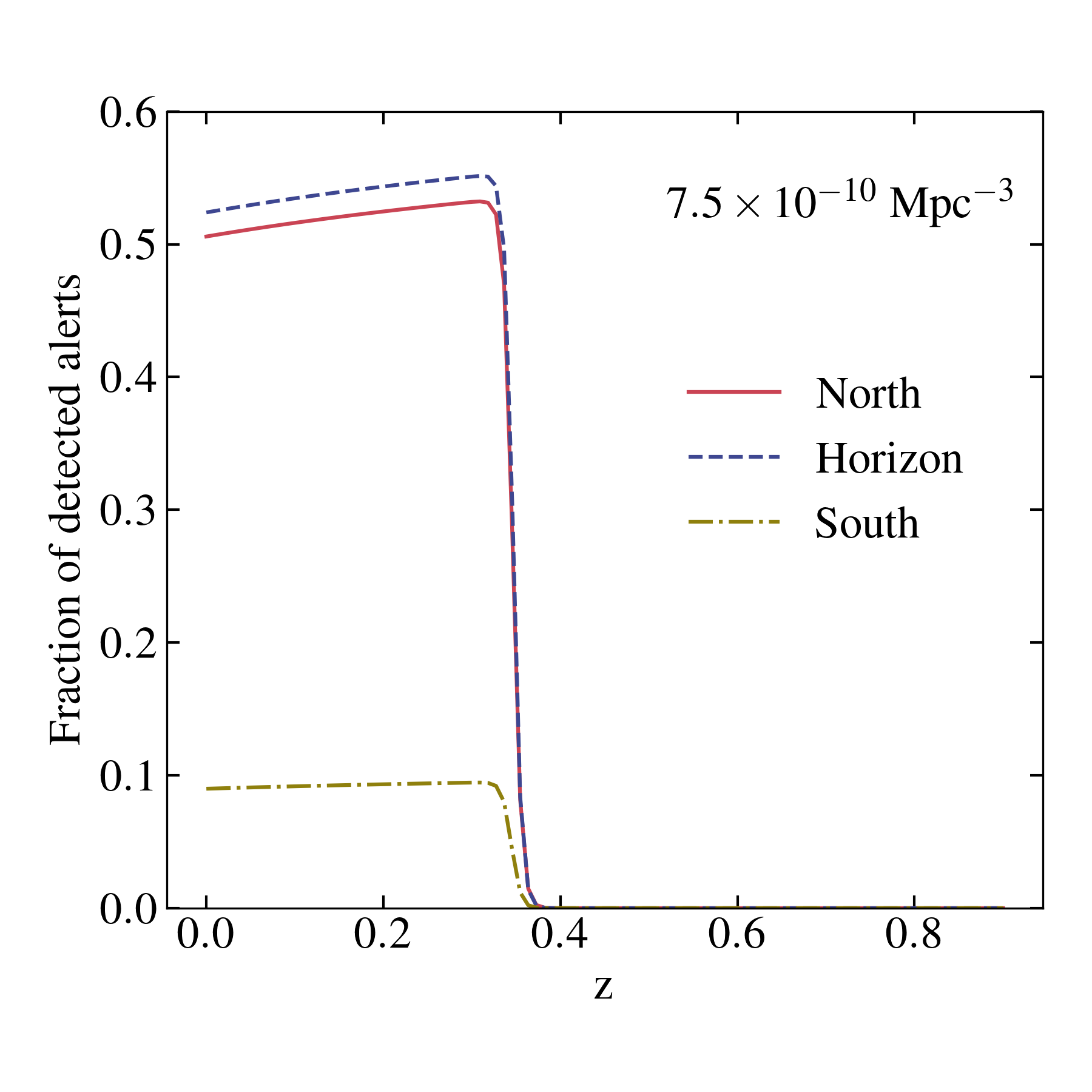}
    \includegraphics[width=0.3\textwidth]{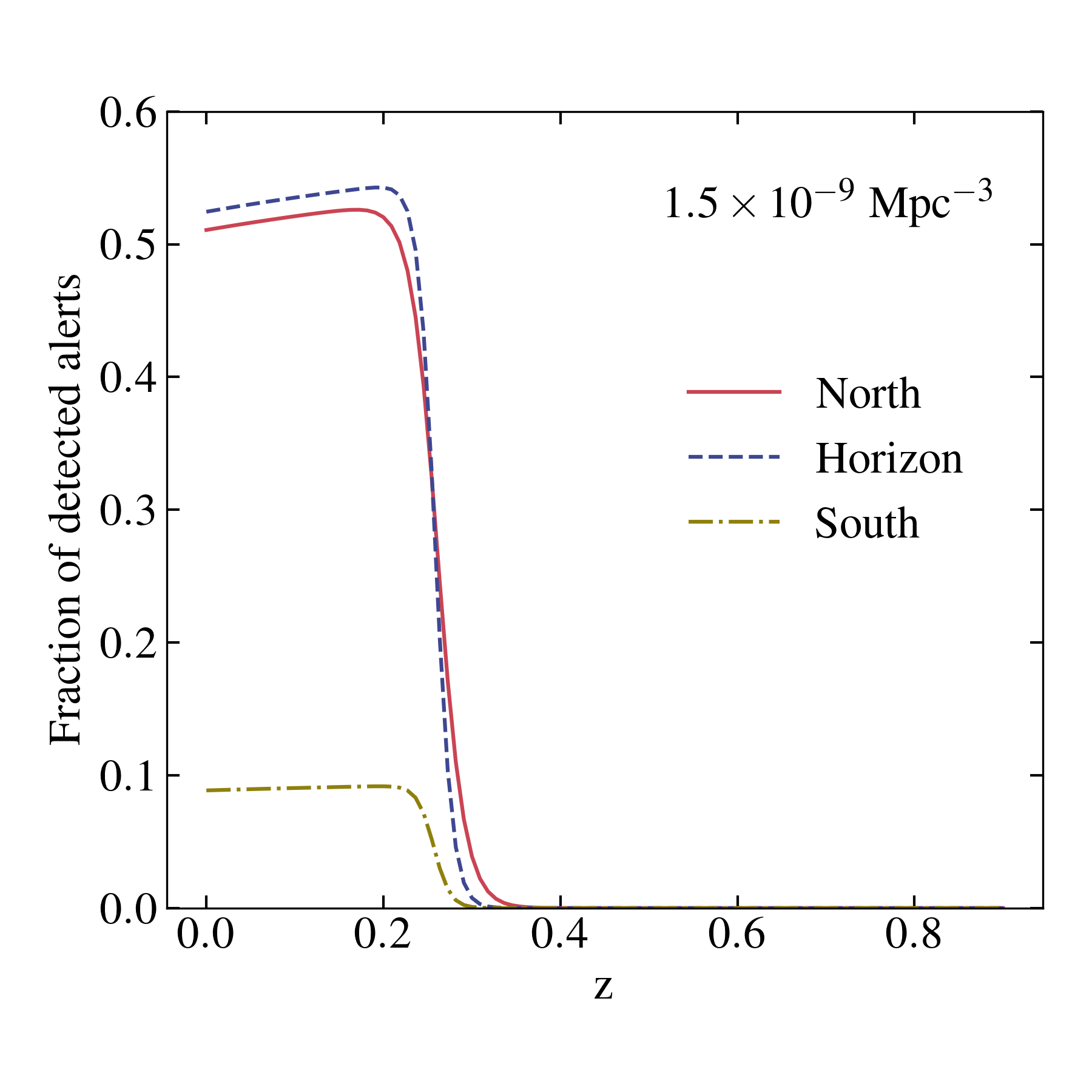}
    \caption{Fraction of detected alerts as a function of the redshift, accounting for observability and flare duration constrains. We show the fraction of detected alerts for each set of simulated alerts corresponding to a local source density of $1.5\times 10^{-10}$ Mpc$^{-3}$ (left panel), $7.5\times 10^{-10}$ Mpc$^{-3}$ (middle panel), and $1.5\times 10^{-9}$ Mpc$^{-3}$ (right panel). In each panel we separate the results for the three regions of declination: northern hemisphere (red solid), horizon (blue dashed) and southern hemisphere (golden dotted-dashed). The decrease with redshift, due to EBL absorption, is discussed in the main text.}
    \label{fig:redshiftreach}
\end{figure}

Using the procedure above, we find that there are two critical factors determining the rate of detected alerts, namely the redshift and the declination of the alerts. To illustrate this point, we show in Figure~\ref{fig:redshiftreach} the fraction of detected alerts as a function of redshift for each set of simulated alerts (different panels corresponding to different local source densities) and for the three angular regions (different colors). 

This result shows that the alerts can be detected up to a maximum redshift. This is due to EBL absorption, which limits the detectability of high redshift sources. The maximum redshift increases for lower values of the local source density. To understand this, we recall that each of the simulated alerts is already detectable by IceCube, for which at a lower source density more luminous sources will be easier to detect. For an IACT, since they are more luminous, they can be detected up to a higher redshift. The fraction of detected alerts also shows a slightly decreasing trend towards low redshifts. This is caused by the finite duration of the flare and the cosmological time dilation, which make sources at high redshifts more likely to be detected while the flare is still active.

The declination region of the source also influences significantly the result, as the different colored curves show. While the maximum redshift of detection does not depend strongly on the declination, the maximum fraction of detected alerts changes from about $0.55$ for alerts from the horizon and the northern hemisphere to $0.1$ for alerts from the southern hemisphere. This is consistent with the results in Figure~\ref{fig:VisibilityAlerts}, where in the southern hemisphere the probability of having too large a zenith for observation (blue dashed curve) is nearly everywhere 1. As a matter of fact the asymptotic value of the fraction of detected alerts at low redshifts (without accounting for the finite flare duration) can be obtained by averaging the probabilities in Figure~\ref{fig:VisibilityAlerts} in the corresponding region of declination.

Our final result for the detected alert rate is obtained by integrating over the alert redshift. In this way we can obtain the total fraction of detected alerts from each of the declination regions and for each set of simulated alerts corresponding to different source densities. We collect these results in Tab.~\ref{tab:results}, together with the total rate of astrophysical neutrino alerts in each declination region that can be detected by MAGIC. For all values of the local source density we find a rate of alerts detected by MAGIC of 1.01 to 1.55 alerts per year; lower values of the local source density lead to slightly higher values of the detected alert rate because of the higher redshifts which can be detected, as discussed above. 

\begin{table}[t]
    \centering
    \begin{tabular}{l|rrr|rrr|r}
    \hline
            Local source density  & \multicolumn{3}{|c|}{Fraction of detected alerts}  & \multicolumn{3}{|c|}{Detected alert rate (yr$^{-1}$)} & Total alert rate \\
         (Mpc$^{-3}$) & North  & Horizon  & South  & North & Horizon & South &  (yr$^{-1}$) \\
          & ($\delta>30^\circ$) & ($-5^\circ<\delta<30^\circ$) & ($\delta<-5^\circ$) & ($\delta>30^\circ$) & ($-5^\circ<\delta<30^\circ$) & ($\delta<-5^\circ$) & \\
         \hline
        $1.5\times10^{-10}$ & 0.17  & 0.15 & 0.032 & 0.32 & 1.17 & 0.064 & 1.55 \\
        $7.5\times10^{-10}$ & 0.13  & 0.11 & 0.025 & 0.25 & 0.86 & 0.050 & 1.16 \\
        $1.5\times10^{-9}$ & 0.12  & 0.095 & 0.022 & 0.23 & 0.74 & 0.044 & 1.01\\
        \hline
    \end{tabular}
    \caption{Summary of fraction of detected alerts for different declination regions; the fraction is defined as the ratio between detected alerts and gold alerts, including the background ones. The rows correspond to different values of local source density. The last column is the astrophysical alert rate, obtained using the rate of non-background gold alerts reported in Table~\ref{tab:alertrates}.}
    \label{tab:results}
\end{table}


\vskip 1mm

\section{Conclusions}\label{sec:conclusions}

We have shown that following the phenomenological model by \citet{Halzen_2019}, IACTs can identify a flaring blazar associated with a neutrino alert. In this model, neutrino/$\gamma$-ray flares last 82~days in the blazar frame, which motivates to follow up alerts over an extended period. We focus on the observation by IACTs of Gold neutrino alerts by IceCube with a rate of $\sim$11 per year and a probability of 50\% of being astrophysical in origin. Using public IceCube and MAGIC performance information and making conservative assumptions on observation constraints, we have calculated the probability of MAGIC detecting such a flaring blazar. Constraints imposed by the Sun and Moon mean that MAGIC should be able to follow-up $\sim$6 Gold alerts per year, assuming up to a 30-day delay between the alert time and the observation. By considering the possibility of the flare being over by the time the observation is made, we find that MAGIC should detect between 1 and 1.5 associated flaring blazars in coincidence with a Gold alert. In our calculations, we have ignored the effects of the inter-galactic magnetic field. This can affect the normalization of the flux near $\sim$100~GeV. An observed $\gamma$-ray could provide insights into the intergalactic magnetic field as discussed in \citet{Halzen_2019} and \citet{2020ApJ...902L..11B}.

The novelty of the observation strategy presented here lies in the delay between the neutrino alert and the IACT observation, which can be as long as 30 days. It is in contrast with the common IACT strategy of making observations mostly very shortly after neutrino alerts. Our results rely on a very modest observation time of 30 minutes per neutrino alert. If the observation time with MAGIC is expanded to 2 hours, the rate of detections is 1.33 to 1.85 per year, depending on the assumed local density of neutrino flaring blazars. For 30 (120) minutes observations per alert as proposed here, MAGIC would dedicate 3 (12) hours per year following this strategy. Calculations presented here are probably applicable for VERITAS, which has similar sensitivity and latitude as MAGIC. The up to month long delay in observations is well suited to minimize observational interference by the Moon. This delay is probably appropriate even if it is assumed that some flares can last longer than the 82~days, self-frame, that we have used in our calculations. 

The new strategy we propose, observing alerts after a longer time delay than that commonly practiced now, complements current observation plans by IACTs and, if implemented, can result in the discovery of a subclass of blazars that would be responsible for the high-energy diffuse neutrino flux. 

\acknowledgments
We are grateful to Anthony M. Brown, Francis Halzen, Ali Kheirandish, Alberto Rosales De Le\'on and Olga Sergijenko for fruitful discussions.
The work of DFGF was supported by the research grant number 2017W4HA7S “NAT- NET: Neutrino and Astroparticle Theory Network” under the program PRIN 2017 funded by the Italian Ministero dell’Istruzione, dell’Universita‘ e della Ricerca (MIUR), and INFN Iniziativa Specifica TAsP.
KS acknowledge the support of the German BMBF (Verbundforschung Astro- und Astroteilchenphysik), HAP (Helmholtz Alliance for Astroparticle Physics) and the European Union’s Horizon 2020 Programme under the AHEAD2020 project (grant agreement n. 871158). IT and CFT acknowledge support by NSF grant PHY-1913607. 


\bibliography{main.bib}{}
\bibliographystyle{aasjournal}

\end{document}